\documentstyle[12pt,epsf]{article}
\newcommand{\preprint}[1]{\begin{table}[t]  
            \begin{flushright}              
            \begin{large}{#1}\end{large}    
            \end{flushright}                
            \end {table}}                           
\preprint{TAUP-2380-96}
\preprint{IHES/P/96/69}
\tiny\normalsize

\begin{document}

\title{A Comment on The $\frac{A}{4}$ Black Hole Entropy.}
\author{E. Atzmon\thanks{%
atzmon@post.tau.ac.il}  \\ 
Raymond and Beverly Sackler Faculty of Exact Sciences,\\ School of Physics
and Astronomy.\\Tel\ - Aviv University.}
\date{October 16,1996}
\maketitle

\begin{abstract}

Using a simple analysis based on the measurement procedure for a quantized area  we explain the $\frac{1}{4}$ factor in the Bekenstein-Hawking black hole formula $\frac{A}{4}$ for the entropy.  
\end{abstract}
\newpage

\section{Introduction}

The existence of black hole entropy was first inferred by Bekenstein \cite{a,b} from the well known connection between entropy and information. Bekenstein`s reasoning was simple and goes as follows. When a heavy star of mass $M$ undergoes a gravitational 
collapse to form a black hole, it carries within it all information about the internal micro-states of its initial state. When the collapsing star reaches its Schwarzschild radius, the information finds itself hidden by the event horizon. If $m$ is the mass of a typical subatomic particle making the collapsed body, and assuming one bit of information per subatomic particle, then the total information lost down the hole and residing in the horizon is roughly $\frac{2M}{m}$. Using the connection between information and entropy, the entropy associated with the formed black hole is proportional to $\frac{2M}{m}$. Bekenstein argued persuasively \cite{a,b} that the Compton wavelength of the subatomic particle (of mass $m$) should be $\frac{1}{m} \le 2M$ in order to fit into the black hole. This automatically sets a maximum value of the information loss and hence on the entropy, i.e. , $S_{max} \sim 4M^2 \sim Area$. The precise relation $S = \frac{\mbox Area}{4}$ was supplied in Hawking's celebrated paper \cite{c}
. 

We propose in this paper another derivation of the $\frac{1}{4}$ factor in the area law. In a way, our method based on measurement considerations complements Bekenstein's beautiful intuition to yield the precise and the full black hole entropy without delving into the details of field theory on a curved background. 

\section{Discrete measurements and the derivation of the area law}

   In any measurement, of an observable property one uses an appropriately selected measurement device. This is generally done by comparison, with the measurement device possessing inherently the same property which it is designed to measure. Common examples are: time is known to be measured by a watch and length by a ruler. 

In order to get accurate results, it would be better that the units of the property in the measurement device be much smaller than the amount of the property found in the measured object.       

   The question we would like to ask is what happen when the magnitude
   of the property of the measured object is precisely that of the
   elementary quantum in the measurement device. In other words if the
   fundamental unit of length, in both the measured object and the
   ruler, is the Planck length $l_p$ (i.e. $\sqrt{\frac{\hbar G}{c^3}}$\ ), what then is the result of measuring one Planck length in space. I claim that the answer to the last question is TWO, the reason being that one needs two Planck units of the ruler to be able to measure one Planck unit of space. 

The last result follows from the fact that the boundaries of a Planck unit in space cannot be adjusted to the boundaries of a one Planck unit of the ruler. This is due to the following facts: 1. For the observer, who is reading the results on the ruler, to be able to observe a {\large point}, he would need to see with an infinite frequency.
 2. The boundaries of the length-segment are points, therefore are of zero dimension and hence zero measure in the standard mathematical measure theory. Thus,
it is impossible to separate a point out of an uncountable set of points.
3. The boundaries of the length-segment are fluctuating (as for an open string), and hence cannot be considered localized. The last argument, of course, depends on what ``nature'' consists of.    

So, the conclusion up to this point, would appear to indicate that the result of the measurement cannot be read as ``one''. 
Our conclusion that the result of the measurement would read ``two'' is based on the following consideration:

\vspace{5mm}
\epsfxsize=11truecm
\centerline{\epsfbox{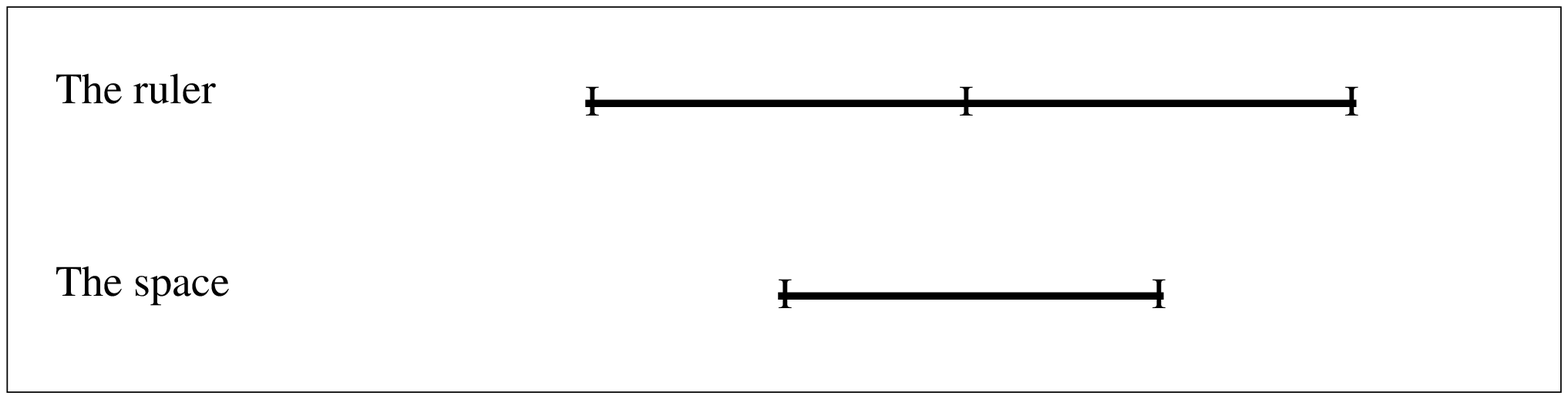}}

\vspace{2mm}
\centerline{\parbox{11truecm}{Figure 1.{\footnotesize The measurement 
process.}}}
\vspace{5mm}

What Fig.1 illustrates is that in order to measure one Planck length with a ruler, one needs essentially to be able to separate between the boundaries of the measured segment. That is to say, that the left boundary of the space segment be covered by one segment of the ruler, whilst the right boundary is covered by another segment. This allows the boundaries to be isolated and distinguished from each other, so that the observer can conclude that the space he has measured has a {\it length}. Since the result of the measurement is being read from the ruler, the result is thus {\it two} Planck length units.\footnote{The same was found in \cite{B.L.S,Atzmon} while using noncommutative geometry to find distances on a one dimensional lattice.}  

In Fig.2 we illustrate the same measurement procedure mentioned above
- except that here the  measurement concerns the area. We assume that
there are Planckons (fundamental units of Planck area -
i.e. $\frac{\hbar G}{c^3}$\ ) which are found both in the measurement device and in the object to be measured. We demonstrate the procedure for both triangles and squares.\\

\vspace{5mm}
\epsfxsize=11truecm
\centerline{\epsfbox{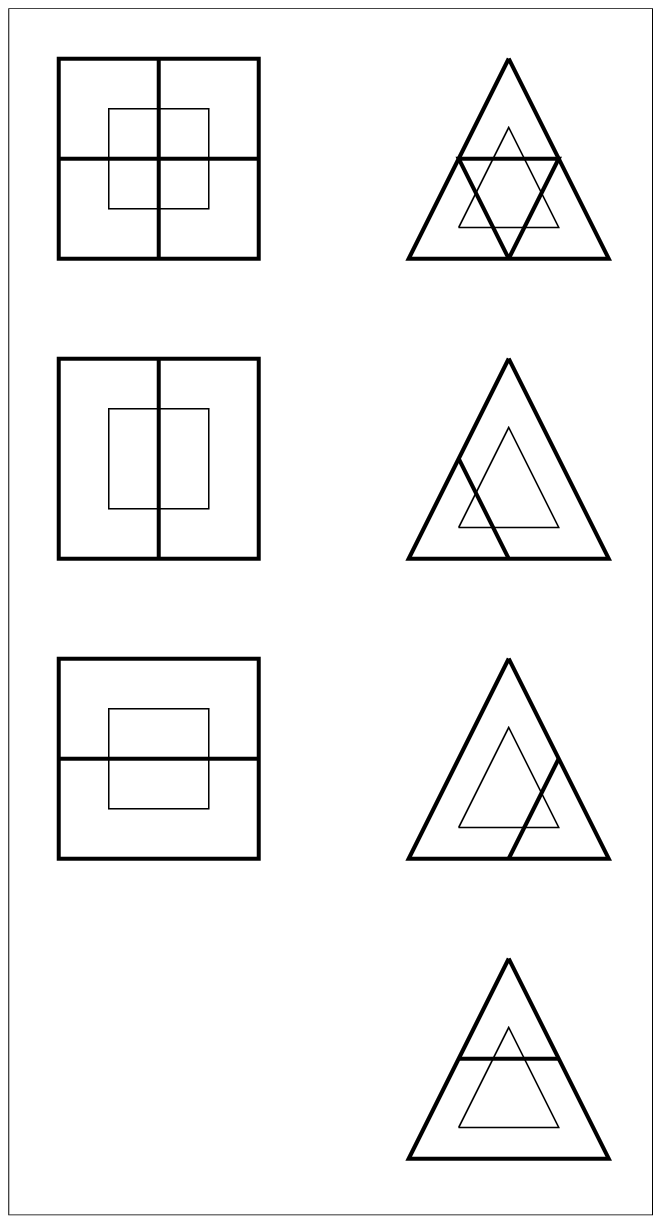}}
\vspace{2mm}
\centerline{\parbox{11truecm}{Figure 2.{\footnotesize The measurement 
process of a one Planckon.}}}
\vspace{5mm}

As one can see, in order to be able to separate the corners of the space plaquettes from each other, one needs four plaquettes of the measurement device. We also demonstrate how the edges can be observed - through the ability of a separating them. The conclusion one draws is that the result of measuring one space Planckon is {\it four}.

Now, if one accepts Bekenstein's interpretation \cite{a,b} - that the bits of information are encoded on the black hole surface, with each Planckon containing one bit of information, then due to the fact that four Planckons of the measurement device are needed in order to detect one space Planckon, we conclude that for a measured black hole area A (as seen by an outside observer while measuring Planckon by Planckon of the black hole surface) the real area is A/4.  

Acknowledgments

I would like to thank Prof. Y. Ne'eman for useful discussions, Dr. N. Hambli for encouraging me to write down this paper and for his help in composing the introduction. I would also like to thank Dr. C. Schweigert for enlightening comments. I am greatful to Prof. A. Connes for inviting me to the IHES, where this work was written, and to the IHES for its warm hospitality.


\begin{thebibliography}{1}
\bibitem{a} J.D. Bekenstein, Lett. Nuovo Cimento 4 (1972) 7371
\bibitem{b} J.D. Bekenstein, Phys. Rev. D 7 (1973) 2333
\bibitem{c} S.W. Hawking, Comm. Math. Phys. 43 (1975) 199 
\bibitem{B.L.S}  G. Bimonte, F. Lizzi, G. Sparano, Distances on a lattice
from noncommutative geometry. Phys. Lett. B 341 (1994) 139-146

\bibitem{Atzmon}  E. Atzmon, Distances on a one-dimensional lattice from
noncommutative geometry, Lett. in Math. Phys. 37 (1996) 341-348

\end{thebibliography}
\end{document}